\documentclass[12pt]{article}
\usepackage{graphicx}
\newcommand{\beq}{\begin{equation}}
\newcommand{\eeq}{\end{equation}}
\newcommand{\bea}{\begin{eqnarray}}
\newcommand{\eea}{\end{eqnarray}}

\begin{document}
\begin{center}
{\Large The Legality of Wind and Altitude Assisted Performances in the Sprints}
\\
\vskip .2 cm 
J.\ R.\ Mureika
\\
{\it Department of Physics} \\
{\it University of Toronto} \\
{\it Toronto, Ontario~~Canada~~M5S 1A7} \\
Email: newt@palmtree.physics.utoronto.ca \\
\end{center}

\vskip .25 cm

\noindent
{\footnotesize
{\bf Abstract} \\
Based on a mathematical simulation which reproduces accurate split and velocity profiles for the 100 and 200 metre 
sprints, the magnitudes of altitude and mixed wind/altitude-assisted performances as compared to their sea-level 
equivalents are presented.  It is shown that altitude-assisted marks for the 200 metre are significantly higher than for the 
100 metre, suggesting that the ``legality'' of such marks perhaps be 
reconsidered.}

\section{Introduction}
According to IAAF regulations, sprint and jump performances for which the measured wind-speed 
exceeds +2.0~m$/$s are deemed illegal, and cannot be ratified for record purposes (IAAF 1998). 
Similarly,  performances which are achieved at altitudes exceeding 1000 metres above sea level are 
noted as ``altitude-assisted'', but unlike their wind-aided counterparts, these can and have qualified 
for record status.  Indeed, the 1968 Olympics saw amazing World Records (WRs) set in the men's 
100~m, 200~m, and Long Jump, thanks in part to the lofty 2250 metre elevation of Mexico City.  
Other examples of such overt assistance include Pietro Mennea's former 200~m WR of 19.72 
seconds, Marion Jones' 1998 clocking of 10.65 s in Johannesburg, Obadele Thompson's wind-
assisted 9.69~s, and Michael Johnson's early 2000-season marks of 19.71~s in the 200 m and 
World Best 30.85~s.

A search of the academic literature reveals a wealth of sources which discuss the impact of wind 
and altitude assistance in the 100 metre sprint. Based both on statistical and theoretical models, the 
general consensus of most researchers is that the maximum legal tail-wind of $+2.0~$m$/$s yields 
roughly a 0.10-0.12 second advantage over still conditions at low altitude.  With no wind, every 
1000 m of elevation will improve a performance by roughly 0.03-0.04 seconds, implying that still 
conditions in Mexico City have about a 0.07 s advantage over their sea-level equivalents.   The 
interested reader is directed to references (Davies 1980, Dapena 1987, Dapena 2000, Linthorne 
1994a, Linthorne 1994b, Mureika 2001a) and citations therein for further information.

Conversely, little attention has been paid to the equivalent corrections in the 200 metres.  
There are multifold reasons why this is perhaps the case, the largest of which being a lack of 
essential wind data for the first half of the race.  The wind gauge is operated only after the first 
competitor has entered the straight, and without a second, perpendicular wind gauge placed at the 
top of the curve, the actual conditions in the first 100 metres remain a mystery.  In a recent article 
(Mureika 2001b), such effects have been studied, accounting for variable wind effects on the curve.  

The model used in this investigation is a direct extension of that presented in (Mureika 2001a), 
whose results are consistent with independent investigations.  The underlying framework of the 
model is a part-mathematical, part-physical force equation of the form

\begin{equation}
F_{\rm net}(t; v, w) = F_{\rm propulsive}(t) - F_{\rm internal}(v;t) - F_{\rm drag}(v,w)~,
\label{eq1}
\end{equation}

Here, $F_{\rm propulsive}(t)$  and $F_{\rm internal}(v;t)$  are functions of  the ``sprinter'', and are dependent on the elapsed 
time $t$ and athlete's resulting velocity $v(t)$.  They are intended to numerically represent both the 
forward driving of the sprinter, as well as any internal variables which govern the overall 
acceleration and speed ({\it e.g.} flexibility, stride rate, fast-twitch rate,  and so forth).  The drag term 
$F_{\rm drag}(v, w; \rho) = 1/2 \, \rho(H) A_d (v(t)-w)^2$ is an external, physical quantity, which is a function of the square 
of the sprinter's relative velocity to the air, the athlete's average drag area Ad (or frontal cross-
sectional area times the drag coefficient, normalized to mass), and the atmospheric density $\rho = \rho(H)$ (dependent on the altitude $H$ of the venue).   
$F_{\rm net}$ is twice-integrated with respect to time to 
obtain the distance traveled as a function of time, $d(t)$, and hence for a suitable choice of input 
parameters, has been shown to effectively and realistically simulate the split/speed profiles of a 100 metre race.

Since the effects of cross-winds ({\it i.e.} winds that are completely perpendicular to the direction of 
motion) are assumed to be negligible, only the forward drag is included.   Hence, for the 200 metre 
simulations, only the component of the wind in the direction of motion are considered (which 
depends on the position of the athlete through on curve).  It is recognized that the influence of a 
strong cross-wind will undoubtedly affect the motion in some fashion, but since these effects are 
currently unknown, they are left for future work. 

This article is not designed to be an expository of the numerical model, but rather a highlight of the 
results which address the ``legality'' of wind and altitude assistance for the 200 m sprint, as 
compared to those in the 100~m.  The interested reader is referred to (Mureika 2001a, Mureika 
2001b) for a complete mathematical and methodological formulation.

\section{Wind and Altitude Effects in the 100~m}

Table~\ref{table1} shows correction estimates for a 10.00 second 100 metre performance run with 0-wind at 
sea level, obtained from the model discussed in (Mureika 2000a).  Corrections are to be interpreted 
as $\Delta t = t_{\rm official} - t_{0,0}$, {\it i.e.} the amount by which the 
0-wind, sea level performance $t_{0,0}$ is adjusted under 
the conditions.  So, negative corrections mean faster official times (and vice versa for positive $\Delta t$).  
Note that a legal-limit wind at high altitude will provide almost a 60\% increase over the same 
conditions at sea-level.  Only for extremely high elevations does the magnitude of the altitude 
assistance alone approach that of the wind, with the minimum altitude-assisted correction being 
about 0.04 seconds.  Only at altitudes at exceeding 2000 m does the assistance begin to approach 
that provided by a low-altitude legal-limit wind.  For a $+1~$m$/$s wind at 2000 m, the theoretical 
assistance is equal to a legal-limit, sea level wind.  These figures are in good agreement with those 
in (Dapena 2000), with showing only mild variations at higher altitudes and wind speeds.

\begin{table}[h]
\begin{center}
{\begin{tabular}{r| r r r r r r }\hline
w (m/s)& 0 m& 500 m& 1000 m& 1500 m& 2000 m& 2500 m\\ \hline
0.0& 0.00& -0.02& -0.04& -0.05& -0.07& -0.08\\
+1.0& -0.05& -0.07& -0.08& -0.10& -0.11& -0.12\\
+2.0& -0.10& -0.11& -0.13& -0.14& -0.15& -0.16\\ \hline \hline
0.0& 0.00& -0.02& -0.04& -0.06& -0.07& -0.09\\
+1.0& -0.07& -0.08& -0.10& -0.11& -0.12& -0.14\\
+2.0& -0.12& -0.14& -0.15& -0.16& -0.17& -0.18\\ \hline
\end{tabular}}
\end{center}
\caption{\footnotesize
Correction estimates (s) for 100 m at varying altitude (Top three rows or men, 
latter three for women), as compared to 10.00 s (11.00 s) performance at sea 
level, 0-wind.}
\label{table1}
\end{table}

An easy to use ``back-of-the-envelope'' formula was presented in 
(Mureika 2001a), which can be 
used to quickly calculate the corresponding corrections of Table~\ref{table1}.
This is

\beq
t_{0,0} \simeq t_{w,H} \left[ 1.03 - 0.03 exp(-0.000125 \cdot H) (1 - w \cdot t_{w,H} /100)^2 \right]~, 
\label{eq2}
\eeq

with $t_{0,0}, t_{w,H}, w$, and $H$ defined as before. Thus, 
100 metre sprint times may be corrected to their 
0-wind, sea level equivalents by inputting only the official time, the wind gauge reading, and the 
altitude of the sporting venue.  Since Equation~\ref{eq2} is easily programmable in most scientific 
calculators and portable computers, it may be used track-side by coaches, officials and the media 
immediately following a race to gauge its overall ``quality''.

\section{Wind and Altitude Effects in the 200~m}

The story, however, is different for the longer sprint.  
Tables~\ref{table2},~\ref{table3} and exemplify the degree of assistance which wind and altitude provide for World Class men and women's 
performances (20.00 s and 22.00 s).  The estimates assume a race run around a 
curve of radius equivalent to about lane 4 of a standard IAAF track, implying 
the distance run around the curve is 115.6~m, and 84.4~m on the 
straight.  The model Equations~\ref{eq1} are modified by the addition of an appropriate ``damping factor'' 
to the propulsive forces (a function of the velocity and the lane's radius).  When coupled with the 
effects of wind, the amplitude of the altitude corrections escalates.  While the absolute value of the 
corrections may not be known at this  point, it is the magnitude of these estimates to which this 
research note draws attention.  

The data presented in Tables~\ref{table2},~\ref{table3} assume that the wind is entirely in the direction measured by the 
gauge.  In this case, the athlete initially faces a head-wind out of the blocks, which gradually 
subsides and increases to its maximum value as the sprinter rounds the bend.   A straight wind of  
$+2.0~$m$/$s adjusts the overall 200~m time by $-0.12~$s for men ($-0.14~$s 
for women), slightly more than 
the correction for the 100~m under similar conditions.  However, the difference between the two 
race corrections quickly grows for increasing wind-speed and altitude.  
In fact, the pure altitude 
effects at the minimum 1000 m elevation are found to be equivalent to that provided by a 
legal-limit wind in the 100 m.  Furthermore, the combined wind and altitude effects could become as 
high as 0.25-0.30 seconds for extreme elevations ($H > 2000$~m).

\begin{table}[h]
\begin{center}
{\begin{tabular}{r| r r r r r r }\hline
w (m/s)& 0 m& 500 m& 1000 m& 1500 m& 2000 m& 2500 m\\ \hline
0.0& 0.0& -0.05& -0.10& -0.15& -0.20& -0.24\\
+1.0& -0.06& -0.11& -0.16& -0.20& -0.24& -0.28\\
+2.0& -0.12& -0.16& -0.20& -0.25& -0.28& -0.32\\ \hline \hline
0.0& 0.0& -0.06& -0.11& -0.16& -0.21& -0.26\\
+1.0& -0.08& -0.16& -0.18& -0.23& -0.27& -0.31\\
+2.0& -0.14& -0.19& -0.23& -0.28& -0.32& -0.35\\ \hline
\end{tabular}}
\end{center}
\caption{\footnotesize
Men's and Women's correction estimates (s) for 200~m at varying altitudes,  as 
compared to 20.00~s (22.00~s) performance at sea level, 0-wind.  
The wind direction is assumed to be completely in the direction of the 
gauge (100 metre straight).}
\label{table2}
\end{table}

Note that the 100 m splits do not significantly change for the wind conditions considered.  Up to  
about 1000 m altitude, the head-wind equivalent conditions in the early part of the race actually  
serve to slow the splits from their 0-wind, sea level equivalent.  Even at high elevations, the 
splits are not significantly affected, being corrected by only -0.05 s at the most.  The  split 
corrections for the 0-wind condition are essentially identical to those for the 100 m, since the 
adjustments depend on the velocity profile over the distance, and are not affected by the curve.

\begin{table}[h]
\begin{center}
{\begin{tabular}{r| r r r r r r }\hline
w (m/s)& 0 m& 500 m& 1000 m& 1500 m& 2000 m& 2500 m\\ \hline
0.0& 0.00& -0.02& -0.03& -0.05& -0.06& -0.08\\
+1.0 & +0.02& +0.00& -0.02& -0.04& -0.05& -0.07\\
+2.0 & +0.03& +0.01& -0.01& -0.02& -0.04& -0.05\\ \hline
\end{tabular}}
\end{center}
\caption{\footnotesize
Correction estimates (s) for 100~m splits of men's (20.00~s) 200~m race.  
0-wind split is approximately 10.25~s including reaction.
}
\label{table3}
\end{table}

Using the corrections of Table~\ref{table2}, one can obtain ``first-order''
adjustments of some key 200~m performances.  For example, Pietro Mennea's 
WR of 19.72 seconds run in Mexico City with a $+1.8~$m$/$s wind would be 
corrected by approximately 0.31~seconds, yielding a 0-wind, sea-level 
equivalent of 20.03~s.  This is essentially equivalent to his low-altitude 
bests, {\it e.g.} (20.01~s; +0.0~m$/$s) in Rome (08 Aug 1980).  Similarly, 
Michael Johnson's 19.71~s ($+1.8~$m$/$s) in Pietersburg 
(approximately 1200~m) would roughly adjust to a mid-19.9~s.  This is also 
quite consistent with his low-altitude bests of 2000, all of which 
clustered around 19.90~s. Tables\ref{table4},~\ref{table5} show the corrected 
top-5 all-time performances for men and women, as well as 
the re-ranked top-5 performances. 

It should be stressed that the calculations herein were performed for an performance around a curve 
of equivalent radius to lane 4 (and appropriate stagger).  The correction estimates for a straight 
wind will actually vary by several hundredths of a second from lane 1 to 8.  For a tail-wind, there 
will be minimal assistance provided in lane 1, and maximal in lane 8.   Thus, lane 4 is selected as
the ``standard'' for conversion.

\begin{table}[h]
{\begin{tabular}{c|c r|c r}\hline
$t_{w,H}$ (w)& $t_{0,0}$ (s)& Athlete& Venue (altitude)& Date\\ \hline
19.32 (+0.4)& 19.38& Michael Johnson  USA& Atlanta, USA (350 m)& 96/08/01\\
19.66 (+1.7)& 19.79& Michael Johnson  USA& Atlanta, USA& 96/06/23\\
19.68 (+0.4)& 19.74& Frank Fredericks  NAM& Atlanta, USA& 96/08/01\\
19.72 (+1.8)& 20.03& Pietro Mennea  ITA& C. de Mexico, MEX (2250 m)& 79/09/12\\
19.73 (-0.2)& 19.73& Michael Marsh  USA& Barcelona, ESP (100 m)& 92/08/05\\
19.75 (+1.5)& 19.86& Carl Lewis  USA& Indianapolis, USA (200 m)& 83/06/19\\ \hline \hline
21.34 (+1.3)& 21.45& Florence Griffith-Joyner  USA& Seoul, SKR (100 m)& 88/09/29\\
21.56 (+1.7)& 21.69& Florence Griffith-Joyner  USA& Seoul, SKR  & 88/09/29\\
21.62 (-0.6) & 21.76& Marion Jones  USA& Johannesburg, SA (1800m)& 98/09/11\\
21.64 (+0.8)& 21.71& Merlene Ottey  JAM& Bruxelles, BEL (50 m)& 91/09/13\\
21.71 (-0.8)& 21.67& Heike Drechsler  GER& Stuttgart, GER (250 m)& 86/08/29\\
21.72 (-0.1)& 21.73& Gwen Torrence  USA& Barcelona, ESP& 92/08/05\\ \hline
\end{tabular}}
\caption{\footnotesize
Official top 5 all-time rankings for men and women, showing 0-wind, 
sea-level equivalents ($t_{0,0}$).  Best-per-athlete (excluding WR).  
Altitudes are assumed correct to within 50~m.  Races are 
assumed run in lane 4.
}
\label{table4}
\end{table}

\begin{table}[h]
{\begin{tabular}{c|c r|c r}\hline
$t_{0,0}$ (w)& $t_{w,H}$ (s)& Athlete& Venue (altitude)& Date\\ \hline
19.38& 19.32 (+0.4)& Michael Johnson  USA& Atlanta, USA& 96/08/01\\
19.72& 20.01 (-3.4)& Michael Johnson  USA& Tokyo, JPN (0 m)& 91/08/27\\
19.73& 19.73 (-0.2)& Michael Marsh  USA& Barcelona, ESP& 92/08/05\\
19.74& 19.68 (+0.4)& Frank Fredericks  NAM& Atlanta, USA& 96/08/01\\
19.75& 19.80 (-0.9)& Carl Lewis  USA& Los Angeles, USA (100 m)& 84/08/08\\
19.83& 19.61 (+4.0)& Leroy Burrell  USA& College Station, USA (300 m)& 90/05/19\\ \hline \hline
21.45& 21.34 (+1.3)& F. Griffith-Joyner   USA& Seoul, SKR& 88/09/29\\ 
21.62& 21.66 (-1.0)& Merlene Ottey  JAM& Zurich, SWI (400 m)& 90/08/15\\
21.67& 21.71 (-0.8)& Heike Drechsler  GER& Stuttgart, GER& 86/08/29\\
21.69& 21.56 (+1.7)& F. Griffith-Joyner  USA& Seoul, SKR& 88/09/29\\
21.73& 21.75 (-0.1)& Juliet Cuthbert  JAM& Barcelona, ESP& 92/08/05\\
21.73& 21.72 (-0.1)& Gwen Torrence  USA& Barcelona, ESP& 92/08/05\\ \hline
\end{tabular}}
\caption{\footnotesize
Corrected top 5 all-time rankings for men and women.  ($t_{0,0}$).  Best-per-athlete 
(excluding WR). Altitudes are assumed correct to within 50~m. Races are assumed run in lane 4.}
\label{table5}
\end{table}
 
%\begin{figure}[h]
%\begin{center} \leavevmode
%\includegraphics[width=0.5\textwidth]{windfig2.ps}
%\end{center} \caption{
%Arbitrary cross-winds blowing at angle $\theta$.
%}
%\label{figure1}
%\end{figure}
 
As previously mentioned, the lack of wind condition information over the first half of the 200~m 
race ultimately prevents completely accurate correction estimates.  For a wind w blowing at angle $\theta$ 
to the straight, the gauge reads $w \, {\rm cos}\, \theta$.  
An angle $\theta = 0$ corresponds to a wind 
purely down the straight, with the value increasing in the counterclockwise direction (such that $\theta < 0$ 
will provide a tail-wind assistance around the bend, and $\theta > 0$
a head-wind).  
Preliminary results 
(see Mureika 2000b) indicate that there is an extremely wide range of variations in corrections for 
200 metre performances apparently run under the ``same'' wind conditions 
(as read by the gauge).  
In fact, for a raw 20.00~s race, the correction differential between 
effective tail-winds and head-winds on the curve could be up to 
0.3~s for lower altitudes.  At high altitudes, this differential could 
exceed 0.6~s!

In 1990, Leroy Burrell ran the fastest-ever wind-assisted 200~m, a startling 
time of 19.61~s (+4.0~m$/$s) (see Table~\ref{table5}).   
Application of the straight-wind corrections would give a 0-wind, sea-level 
time of 19.83~s, much faster than his legal sea-level best of 20.12~s 
($-0.8~$m$/$s, New Orleans; 20.06~s corrected).  However, 
a wind in excess of 5~m$/$s blowing at an angle of roughly -40~degrees would 
produce the proper gauge reading, and assist the performance by up to 0.4~s, 
much more consistent with Burrell's previous marks.  Also, if this race had 
been run in a higher lane than 4, the correction 
would be larger (but still faster than his previous bests, if the wind is 
assumed purely in the 100~m direction).

If these correction estimates are accurate, then the suggestion is put forth to the IAAF that the status 
of higher (but legal) wind and altitude-assisted 200 metre marks be reconsidered.  The degree 
of variation from differing wind angles would also affect the IAAF Top Performance lists 
introduced this year, as well as similar scoring tables which account for accompanying wind 
speeds.  Equivalently, these types of corrections could be used to ``rate'' the quality sprints run 
under varying conditions.  A wind gauge placed at the top of the curve could help to shed light on 
these effects, as well as assist in the evaluation and comparison of race performances.

\pagebreak
\vskip .3cm
\noindent{\Large {\bf Acknowledgements}} \\
This work was supported in part by grants from the Walter C. Sumner Foundation, 
and the Natural Sciences and Engineering Research Council of Canada (NSERC).

\vskip .3cm
\noindent{\Large {\bf References}} \\

\noindent(IAAF 1998)  {\it Official 1998/1999 Handbook}, International Amateur Athletics Federation, Monaco (1998) \\

\noindent(Davies 1980) C.\ T.\ M.\ Davies, ``Effects of wind assistance and resistance on the forward motion of a runner'', J.\ Appl.\ Physio.\ {\bf 48}, 702-709 (1980)\\

\noindent( Dapena 1987) J.\ Dapena and M.\ Feltner,  ``Effects of wind and altitude on the times of 100-meter sprint races'', Int.\ J.\ Sport Biomech.\ {\bf 3}, 6-39 (1987)\\

\noindent(Dapena 2000) J.\ Dapena, in {\it The Big Green Book}, Track and Field News Press (2000)\\

\noindent(Linthorne 1994a) N.\ P.\ Linthorne, ``The effect of wind on 100~m sprint times'', J.\ App.\ Biomech.\ {\bf 10}, 110-131 (1994)\\

\noindent(Linthorne 1994b) N.\ P.\ Linthorne, ``Wind and altitude assistance in the 100-m sprint'', {\it Proc.\ 8th Bienn.\ Conf.\, Can.\ Soc.\ Biomech.}, W.\ Herzog, B.\ Nigg and T.\ van den Bogert (Editors), 68-69 (1994)\\

\noindent(Mureika 2001a)  J.\ R.\ Mureika, ``A Realistic Quasi-physical Model of  the 100 Metre Dash'', to appear, {\it Canadian Journal of Physics} (2001)\\

\noindent(Mureika 2001b) J.\ R.\ Mureika, ``Modeling Wind and Altitude Effects in the 200 Metre Sprint'', in preparation  (2001)\\

\end{document}